 \documentstyle[12pt]{article}
\begin{document}
\begin{flushright}
  hep-ph/9503403
\end{flushright}
\title{Wavelet Spectra of JACEE Events}
\author{Naomichi Suzuki, Minoru Biyajima ${}^*$ and
Akinori Ohsawa ${}^{**}$ \\
 Matsusho Gakuen Junior College, Matsumoto 390-12     \\
 ${}^{*}$ Department of Physics, Facalty of Liberal Arts,
          Shinshu University, \\
          Matsumoto 390        \\
 ${}^{**}$ Institute for Cosmic Ray Research,
University of Tokyo,  \\
           Tanashi 188    \\}
\maketitle
\begin{abstract}
Pseudo-rapidity distributions of two high multiplicity events
Ca-C and Si-AgBr observed by the JACEE are analysed by the
wavelet transform.   Wavelet spectra of those events are
calculated and compared with the simulation calculations.
The wavelet spectrum of Ca-C event somewhat resembles to that
simulated with the uniform random numbers.   That of Si-AgBr
event is not reproduced by simulation calculations with Poisson
random numbers, uniform random numbers, or a p-model.
\end{abstract}

{\it Introduction}.
 In high energy ncleus-nucleus (AA) collisions, number density of
secondary particles in the rapidity space become very high, and
studies of number density fluctuations in the rapidity space is
expected to reveal new features of multiparticle production
mechanisms.  We have analysed pseudo-rapidity distributions in
AA collisions including JACEE events, using Higuchi's method
(a kind of length method), the auto-regressive model \cite{suzu91},
and the fast Fourier transform \cite{biya91}. In order to classify
high multiplicity events by those pseudo-rapidity distribution
patterns, the fractal dimensions of the distributions
(box dimensions) are estimated with both methods.

In this paper, the pseudo-rapidity distributions of the two
JACEE events, Ca-C and Si-AgBr are analysed by the wavelet
transform \cite{daub88},\cite{stra89},\cite{newl93},\cite{mori88}.
  Any function (data) can be expanded by self-similar wavepackets
in this scheme.  Therefore characteristics of local fluctuations
can be extracted from the data. Wavelet spectra of the two events
are calculated and are compared with simulation calculations.  In
the second paragraph, the wavelet transform is concisely introduced.
In the third one, wavelet spectra of the two events are calculated.
Those are compare with the simulation calculations in the next
paragraph. The final paragraph is devoted to the concluding
remarks.

\vspace{5mm}

{\it Wavelet transform}.
 Wavelets are constructed from dilation and translation of a
scaling function.  The scaling function $\phi(x)$ is constructed
by an iteration equation,
 \begin{equation}
         \phi_i(x)= \sum_{k=0}^{N-1} c_{k}\phi_{i-1}(2x-k)
            \quad i=1,2,\cdots,         \label{eq:wvlt1}
 \end{equation}
from a primary scaling function $\phi_0(x)$, where N is a even
number, and $c_k$ $\,(k=0,1,\cdots,N-1)$ are constants.   The
iteration is continued until $\phi_{i}(x)$ becomes
indistinguishable
from $\phi_{i-1}(x)$, and $\phi(x)$ is defined by $\phi_i(x)$.
The primary scaling function $\phi_{0}(x)$ is taken as
 \[  \phi_{0}(x) = \left\{
              \begin{array}{rl}
               1  & \qquad \mbox{for}\quad 0\leq x < 1,  \\
               0  & \qquad \mbox{otherwise}.
               \end{array} \right.
 \]
The mother wavelet $W(x)$ is given by
  \begin{eqnarray}
      W(x) = \sum_{k=0}^{N-1} (-1)^k c_{k}\, \phi(2x+k-N+1),
\label{eq:wvlt1a}
  \end{eqnarray}
and the j-th level wavelet $(j=0,1,\cdots)$ is defined as
  \begin{equation}
       \psi_{j,k}(x) = 2^{\frac{j}{2}} W(2^{j}x-k),   \qquad
                       k=0,1,\cdots,2^{j}-1.  \label{eq:wvlt2}
  \end{equation}
Coefficients $c_k\,(k=0,1,\cdots,N-1)$ should be determined so
that the wavelets and the scaling function satisfy the following
orthogonal conditions,
  \begin{eqnarray}
      \int \phi(x)\,\phi(x) dx &=& 1,    \nonumber \\
  \int \phi(x)\,\psi_{j,k}(x) dx &=& 0,    \label{eq:wvlt3} \\
  \int \psi_{j,k}(x)\,\psi_{r,s}(x) dx &=& \delta_{jr}\delta_{ks}.
                                                \nonumber
  \end{eqnarray}
If N=4, the coefficients are given by
 \begin{eqnarray}
      c_0 = \frac{1+\sqrt{3}}{4}, \qquad\qquad
      c_1 = \frac{3+\sqrt{3}}{4},    \nonumber \\
                                         \label{eq:wvlt4} \\
      c_2 = \frac{3-\sqrt{3}}{4}, \qquad\qquad
      c_3 = \frac{1-\sqrt{3}}{4}.    \nonumber
 \end{eqnarray}

Arbitrary function $f(x)$ defined in the region $0 \le x < 1$
can be expanded as%
 \begin{eqnarray}
       f &=& f^\phi + \sum_{j=0} f^{(j)},   \nonumber   \\
    f^\phi &=& a_{0}\phi(x),         \label{eq:wvlt5}  \\
   f^{(j)} &=& \sum_{k=0}^{2^{j}-1} \alpha_{j,k} \psi_{j,k}(x).
\nonumber
  \end{eqnarray}
By the use of Eqs.(\ref{eq:wvlt3}) and (\ref{eq:wvlt4}),
We have a relation,
 \begin{equation}
     \int f(x)^{2} dx = a_{0}^{2} + \sum_{j=0}
E_{j}, \label{eq:wvlt6}
 \end{equation}
where $E_j$ denotes the j-th level wavelet
spectrum \cite{yama91},
  \begin{eqnarray}
      E_{j} = \sum_{k=0}^{2^{j}-1} \alpha_{j,k}^{2},
       \qquad\quad j=0,1,\cdots.    \label{eq:wvlt7}
  \end{eqnarray}
Equation (\ref{eq:wvlt6}) corresponds to the Parseval's
equation in the Fourier transform.

Hereafter we consider the case that the function $f(x)$
defined in the region $0\le x <1$ is given by descrete value,
  \[   f_i = f(x_1 + \Delta \cdot (i-1)),
\quad i=1,2,\cdots,2^r. \]
We assume that $f(x)$ is periodic outside the defined region.
As the data points are $2^r$, $f(x)$ is expanded by finite terms.
Summation on the right hand side (RHS) of the Eq.(\ref{eq:wvlt4})
runs from $j=0$ to $j=r-1$.

By the use of column matrices,
 \begin{eqnarray*}
         f = \left( \begin{array}{c}
                   f_{1}         \\
                   f_{2}         \\
                   \cdots        \\
                   f_{2^r}
                     \end{array} \right),  \qquad\quad
        A_{j} = \left( \begin{array}{c}
                        \alpha_{j,0}  \\
                        \alpha_{j,1}  \\
                        \cdots        \\
                        \alpha_{j,2^{j}-1}
                       \end{array} \right),    \qquad j=0,1,\cdots,
 \end{eqnarray*}
Eq.(\ref{eq:wvlt5}) is expressed as
  \begin{eqnarray}
    f &=& f^\phi + \sum_{j=0}^{r-1} f^{(j)},   \nonumber  \\
   {}^{t}f^\phi &=& 2^{\frac{r}{2}}\,a_{0}L_{1}L_{2}\cdots L_{r},
\nonumber \\
    {}^{t}f^{(j-1)} &=& 2^{\frac{r}{2}}\,\,
              {}^{t}A_{j-1}H_{j}L_{j+1}\cdots L_{r},  \qquad
               j=1,2,\cdots,r-1,   \label{eq:wvlt8} \\
    {}^{t}f^{(r-1)} &=& 2^{\frac{r}{2}}\,\,
{}^{t}A_{r-1} H_{r}\nonumber,
  \end{eqnarray}
where $L_j$ and $H_j$ are $2^{j-1}\times 2^{j}$ matrices.
The ${\it i}-$th row of $L_j$ has $2(i-1)$'s zeroes from the
first to the $2(i-1)$ -th elements, and is expressed as,
 \begin{eqnarray*}
      \left( \begin{array}{cccccccccc}
    0 & \cdots & 0 & l_{2(i-1)+1} & l_{2(i-1)+2} & \cdots &
          l_{2(i-1)+N} & 0 &  \cdots & 0
          \end{array} \right)          \nonumber  \\
      =  \frac{1}{\sqrt{2}}
      \left( \begin{array}{cccccccccc}
       0 & \cdots & 0 & c_{0} & c_{1} & \cdots &
          c_{N-1} & 0 &  \cdots & 0
          \end{array} \right).
 \end{eqnarray*}
The ${\it i}-$th row of $H_j$ also have $2(i-1)$'s zeroes from
the first to $2(i-1)$-th elements and is written as
 \begin{eqnarray*}
    \left( \begin{array}{cccccccccc}
    0 & \cdots & 0 & h_{2(i-1)+1} & h_{2(i-1)+2} & \cdots &
     h_{2(i-1)+N} & 0 &  \cdots & 0
     \end{array} \right)   \nonumber      \\
     = \frac{1}{\sqrt{2}}
     \left( \begin{array}{cccccccccc}
      0 & \cdots & 0 & c_{N-1} & -c_{N-2} & \cdots &
         c_{0} & 0 &  \cdots & 0
         \end{array} \right).
 \end{eqnarray*}
If $2(i-1)+k > 2^j $ in $L_j$ ( or $H_j$ ), element
$l_{2(i-1)+k}$ ( or $h_{2(i-1)+k}$ ) is added to the
$(2(i-1)+k-2^j)$ -th element in each row.

{}From Eq.(\ref{eq:wvlt4}), matrices $L_j$ and $H_j$ satisfy
the following conditions,
 \begin{eqnarray}
      L_j\,^{t}L_j = I,  \qquad\qquad  L_j\,^{t}H_j = 0,
\nonumber \\
                                    \label{eq:wvlt9}   \\
      H_j\,^{t}L_j = 0,  \qquad\qquad
H_j\,^{t}H_j = I, \nonumber
 \end{eqnarray}
where ${}^tL_j$ denotes the transpose of matrix $L_j$, and
$I$ is the $2^{j-1}-$th order unit matrix.
Then, the wavelet coefficients are written as
  \begin{eqnarray}
      a_{0} &=& 2^{-\frac{r}{2}}L_{1}L_{2}\cdots L_{r}f,
\nonumber   \\
    A_{j-1} &=& 2^{-\frac{r}{2}}H_{j}L_{j+1}\cdots L_{r}f,
           \quad j=1,2,\cdots,r-1, \label{eq:wvlt10} \\
        A_{r-1} &=& 2^{-\frac{r}{2}}H_{r}f.     \nonumber
  \end{eqnarray}
As the summation on the RHS of Eq.(\ref{eq:wvlt6}) runs from
$j=0$ to $j=r-1$ in this case, the wavelet spectra $E_j$ from
$j=0$ to $j=r-1$ are obtained.

\vspace{5mm}

{\it Wavelet spectra of the data}.
 Pseudo-rapidity $\eta$ distributions of Ca-C and Si-AgBr events
are shown in Fig.1a and b, respectively. From each distribution,
we subtract the background distribution \cite{taka84},
 \begin{eqnarray}
    f_{0}(\eta) = A \left[ (1-e^{-Y-\eta})(1-e^{-y+\eta})
\right]^B, \quad
      -Y \le \eta \le Y,    \label{eq:wvlt11}    \\
       A=81,\,  B=5.4, Y=7.0  \qquad  {\rm for\, Ca-C},
\nonumber \\
          A=184, B=8.1,  Y=5.5  \qquad  {\rm for\, Si-AgBr}.
\nonumber
 \end{eqnarray}

\vspace{3mm}
Pseudo-rapidty distributions of Ca-C and Si-AgBr events are given
with the bin size $\Delta\eta=0.1$, and the number n of the data
points in each event is in the range $ 2^6 < n < 2^7$. Those used
in the analysis should be $2^r$. Therefore 64 points $(r=6)$ are
used for both events; $-3.0 \le \eta < 3.4 $ for Ca-C, and $ -2.7
\le \eta < 3.7$ for Si-AgBr.
Fig 2a and b show the distributions of the two events, where the
backgrounds are subtracted. The standard deviation $\sigma$
of the distribution shown in Fig.2a is 2.71, and that of
fig.2b is 4.03.

In Fig.3a and 3b, the wavelet spectra $E_j\,(j=0,1,\cdots,5)$ of
Ca-C and Si-AgBr are shown, respectively.  The black circles show
the wavelet spectra of the distributions where backgrounds are
subtracted. The blank circles show that of the raw distributions
( without subtraction of the backgrounds).  The results are
different at the 0-th and the first level, but those from j=2 to
j=5 are almost the same.  The wavelet spectrum of Ca-C event
increases linearly in semi-logarithmic scale from j=2 to j=5.
That of Si-AgBr is approximately flat in the same range.

\vspace{5mm}

{\it Comparison with simulation calculations}.
The wavelet spectra of the two events are compared with
the following simulation calculations;

 \begin{itemize}
 \item[(i)] Poisson random numbers with a mean $\mu$.  The
standard deviation $\sigma$ of the Poisson distribution with
the mean $\mu$ is given by $\sigma=\sqrt{\mu}$.

\item[(ii)] uniform random numbers from 0 to $a$.  The mean
$\mu$ is given by,
 \[ \mu = \frac{1}{a}\int_0^a\,x\,dx = \frac{a}{2}. \]
The variance $\sigma^2$ is given by
  \[  \sigma^2 =  \frac{1}{a}\int_0^a(x-\mu)^2\,dx =
\frac{a^2}{12}.  \]
Then the parameter $a$ is expressed by the standard deviation as
  \[  a = 2\sqrt{3}\,\sigma.   \]

\item[(iii)] a p-model\cite{mene87} with a fraction $p_a  \,
(p_b=1-p_a)$ and an initial 'energy' (or particle number) $E_0$.

In the p-model, the initial energy $E_0$ in a rapidity
interval $\Delta y$ is devided into $p_a\,E_0$ and $p_b\,E_0$
at the first step, as the interval $\Delta y$ splits into two
sub-intervals with equal width. $p_a\,E_0$ is regarded as the
energy in one interval, and $p_b\,E_0$ is that in the other
interval.  At the subsequent step, each energy is devided into
two in the same way, as each sub-interval splits into two.
After n steps, the initial energy $E_0$ is devided into $2^n$
terms; energy $p_a^j p_b^{n-j} E_0 \quad (j=0,1,\cdots,n)$
appears ${}_nC_j$ times. Those are energies in $2^n$ ordered
sub-intervals of $\Delta y$ with the same width. The mean $\mu$
 and the variance $\sigma^2$ are given respectively by
  \begin{eqnarray*}
      \mu &=& \frac{1}{2^n}\sum_{j=0}^{n} E_0\,p_a^j\,q_b^{n-j}
          = \frac{E_0}{2^n},    \\
     \sigma^2 &=& \frac{1}{2^n}\sum_{j=0}^{n}
         \left( E_0\,p_a^j\,q_a^{n-j} \right)^2 {}_nC_j - \mu^2
         = \left( \frac{E_0}{2^n} \right)^2
         \left( 2^n \left(p_a^2+q_a^2 \right)^n - 1 \right).
 \end{eqnarray*}
In the simulation calculations, we iterate $r+2 = 8$ steps, and
randomize the ordering of the energies.  Then, we add every four
energies, and have $2^6$ ones, which are compared with the data.
 \end{itemize}

Each simulation calculations are done 1000 events.  The mean
value
is not sensitive to the wavelet spectra, and parameters are
adjusted to reproduce the standard deviation of each
distribution shown in Fig.2a or b.

We calculate the wavelet spectra $E_j^{simu}\quad
(j=0,1,\cdots,5)$
of simulated events and count the number of events
within a value of z,
 \begin{equation}
    z = \sqrt{ \sum_{j=2}^5(\log_{2}E_j - \log_{2}E_j^{simu})^2 }.
                                     \label{eq:wvlt12}
 \end{equation}

In order to generate the random numbers, we use a congruence
method\cite{mori86}. At first, we choose an integer $r_0$,
using the equation
 \begin{eqnarray}
    r_{j} = a\,r_{j-1} + c  \qquad {\rm mod}\,\,m,
\qquad j=1,2,\cdots,
                   \label{eq:wvlt13}\\
     a=1229, \quad c=351750, \quad m=1664501,    \nonumber
 \end{eqnarray}
we generate the random numbers $r_j\,\,(j=1,2,\cdots)$
subsequently.  From those random numbers we get the
uniform random number
   \[  s_j = \frac{r_j}{m},   \quad  0\le s_j <1.   \]

The results for Ca-C and Si-AgBr events are shown in
Table 1a and 1b,
respectively.  Table 1a shows that the wavelet spectrum of
simulated events for Ca-C events have more than 15 $\%$ out
of 1000 simulated events within $z < 0.8$ in all the  three
cases.   Those with uniform random numbers have 20 $\%$ and
is higher than the other two cases.

Contrary to the Ca-C event, simulated events for Si-AgBr
event have scarecely low rate within $z<0.8$ in all the three
cases (See table 1b).  The wavelet spectrum of Si-AgBr event
cannot be reproduced by any of these simulation calculations.

Examples of the simulation calculations with (i) Poisson
random numbers are shown in Fig.4a and b, with (ii) uniform
random numbers in Fig.5a and b, and with (iii) the p-model
in Fig 6a and b.

\vspace{5mm}

{\it Concluding Remarks}.
 Two high multiplicity events Ca-C and Si-AgBr are analysed
by the wavelet transform.  Wavelet spectra of the two events
are calculated and compared with the simulation calculations.
 Wavelet spectrum of the Ca-C event somewhat resembles to the
simulation calculations with uniform random numbers, but that
of the Si-AgBr event can not be reproduced by any of the three
simulation calculations.  Further investigation will be reported
elsewhere that there are any simulation calculations which
reproduce $\eta$ distributins of high multiplicity events.
Our analysis show that observed high multplicity events can be
classified by the wavelet spectra, and what statistical
regularity (or irregularity) would be dominant event by event.

\vspace{1cm}
\noindent{\rm Acknowledgements.}

The authors wishes to thank T. Matsunawa for the financial
support for the workshop held at Institute of Mathematical
Statistics in 1991, and helpful comments and discussions.
Authors are also thankful to I. Dremin, A. Morimoto, H.
Takayasu, and M. Yamada for their correspondence and useful
discussions at the initial stage of this work.  This work
is partially supported by the JSPS Program on Japan-FSU
Scientists Collaboration in 1994.  M.B. is partially
suported by Japanese Grant-in-Aid for Scientific Research
from the Mnistry of Education, Science and Culture
(No. 06640383). N.S. thanks for the financial support
by Matsusho Gakuen Junior College in 1994.

\newpage

\begin{flushleft}
{\large Table  Caption}
\end{flushleft}

 \begin{itemize}
  \item[Table 1] Number of simulated events which satisfy
the condition $z<0.4$, $z<0.6$, or $z<0.8$  for a) Ca-C and
b) Si-AgBr. Each simulation calculations are done 1000
events.
 \end{itemize}

\vspace{0.5cm}

\begin{flushleft}
{\large Figure Captions}
\end{flushleft}

 \begin{itemize}
 \item[Fig.1]  Pseudo-rapidity distributions of
a) Ca-C and b) Si-AgBr events.

 \item[Fig.2]  Distributions used for wavelet analysis. Those
are obtained from $\eta$ distributions where the backgrounds are
 subtracted; a) Ca-C, and b) Si-AgBr.

\item[Fig.3]  Wavelet spectra of a) Ca-C and b) Si-AgBr events.
Black circles are calculated from the $\eta$ distributions
where
the backgrounds are subtracted.  Blanck circles are from $\eta$
distributions ( where backgrounds are not subtracted).

\item[Fig.4]  Wavelet spectra of simulated events with Poisson
random numbers for a) Ca-C  with $r_0=44255$ (z=0.14) and b)
Si-AgBr with $r_0=115383$ (z=0.50).

\item[Fig.5]  Wavelet spectra of simulated events with uniform
random numbers for a) Ca-C $r_0=6634423$ (z=0.25) and b)
Si-AgBr with $r_0=115383$ (z=0.55).

\item[Fig.6]  Wavelet spectra of simulated events with the
p-model  for a) Ca-C with $r_0=1567321$ (z=0.27) and b) Si-AgBr
with $r_0=707048$ (z=0.58).
 \end{itemize}
\newpage
\vspace{2cm}
\begin{center}
\begin{tabular}{lccc}\hline
 {criterion} & {Poisson} & {Uniform} &{p-model} \\
      & {($\mu=7.6$)} & {($a=9.40$)} & {( $p_a=0.36,
E_0=380$)} \\ \hline
    {$z<0.4$} & { 18} & { 25} & { 18} \\
    {$z<0.6$} & { 61} & { 75} & { 65} \\
    {$z<0.8$} & {152} & {197} & {168} \\  \hline
\end{tabular}

\vspace{2mm}
           a

\vspace{1cm}

\begin{tabular}{lccc}\hline
 {criterion} & {Poisson} & {Uniform} & {p-model}  \\
     & {($\mu=17.0$)} & {($a=14.8$)} & {($p_a=0.36, E_0=590$)}
\\ \hline
    {$z<0.4$} & 0 & 0 & { 0} \\
    {$z<0.6$} & 2 & 2 & { 2} \\
    {$z<0.8$} & 7 & 8 & {10} \\  \hline
\end{tabular}

\vspace{2mm}
              b

\vspace{5mm}
 Table 1
\end{center}
\end{document}